\RequirePackage{fix-cm}
\documentclass[smallextended]{svjour3}       
\smartqed  
\usepackage{graphicx}
 \usepackage{mathptmx}      
\usepackage{latexsym}

\begin{document}

\title{EChO Payload electronics architecture and SW design}



\author{M. Focardi         \and
        A. M. Di Giorgio \and M. Farina \and M. Pancrazzi \and R. Ottensamer \and T. L. Lim
         \and S. Pezzuto \and G. Micela\and E. Pace 
}


\institute{\textbf{Mauro Focardi} \at
              INAF - Osservatorio Astrofisico di Arcetri - 50125 Firenze, ITALIA\\
              Largo E. Fermi 5 \\
              Tel.: +39-055-275 5213\\
              Fax: +39-055-275 5252\\
              \email{\emph{mauro@arcetri.astro.it}}           
           \and
           \textbf{A. M. Di Giorgio \and M. Farina \and S. Pezzuto}\at
           INAF - Istituto di Astrofisica e Planetologia Spaziale - 00133 Roma, ITALIA;
           \and
           \textbf{M. Pancrazzi} \at
           INAF - Osservatorio Astrofisico di Arcetri - 50125 Firenze, ITALIA;  
           \and
           \textbf{R. Ottensamer} \at
           Universit\"at Wien - Institut f\"ur Astrophysik T\"urkenschanzstr - 1180 Wien, AUSTRIA;
            \and
           \textbf{T. L. Lim} \at
           RAL Space - Rutherford Appleton Laboratory - OX11 0QX Harwell Oxford, UK; 
            \and
           \textbf{G. Micela} \at
           INAF - Osservatorio Astronomico di Palermo - 90134 Palermo, ITALIA; 
            \and
           \textbf{E. Pace} \at
           Universit\`a degli Studi di Firenze - Dept. Fisica e Astronomia - 50125 Firenze, ITALIA.     
   }

\date{Received: date / Accepted: date}

\maketitle

\begin{abstract}
EChO is a three-modules (VNIR, SWIR, MWIR), highly integrated spectrometer, covering the wavelength range from 0.55 $\mu$m, to 11.0  $\mu$m. The baseline design includes the goal wavelength extension to 0.4 $\mu$m while an optional LWIR module extends the range to the goal wavelength of 16.0 $\mu$m.

An Instrument Control Unit (ICU) is foreseen as the main electronic subsystem interfacing the spacecraft and collecting data from all the payload spectrometers modules. ICU is in charge of two main tasks: the overall payload control (Instrument Control Function) and the housekeepings and scientific data digital processing (Data Processing Function), including the lossless compression prior to store the science data to the Solid State Mass Memory of the Spacecraft. These two main tasks are accomplished thanks to the Payload On Board Software (P-OBSW) running on the ICU CPUs.

\end{abstract}

\section{Introduction}
\label{intro}

The Exoplanets Characterisation Observatory \cite{Tinetti_1} is designed as a dedicated survey mission for the study of the atmospheres of a selected sample of extra-solar planets. This main scientific goal can be achieved by using the transit and eclipse spectroscopy method, to observe exoplanets for which the contrast between planet and host star can be as low as $10^{-5}$, and will be typically of the order of $10^{-4}$. As a consequence, the main aim of the EChOÕs payload instrumentation will be to perform time-resolved spectro-photometry (no angular resolution is required as no imaging is foreseen on FPAs) exploiting the temporal and spectral variations of the signal due to the primary and secondary occultations occurring between the exoplanet and its parent star. Moreover, in order to extract the planet spectral signature and probe the thermal structure and the physical and chemical properties of its atmosphere \cite{Tinetti_2}, it will also be necessary to have a large and simultaneous wavelength coverage to detect as many molecular species as possible and to correct for the contaminating effects due to the stellar photosphere at the same time.

\begin{figure}[!h]
\begin{center}
 \includegraphics[width=10cm]{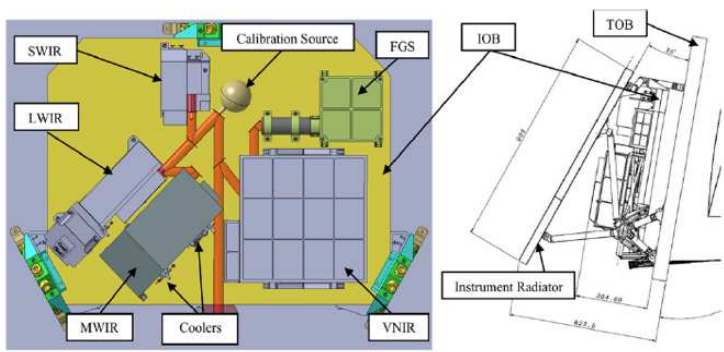}
\caption{EChO Payload optical bench hosting the Science Modules.}
\label{fig:1}       
\end{center}
\end{figure}

To fulfil all these requirements, the EChO scientific payload \cite{EChO_1} has been conceived as a multi-modules highly integrated spectrometer covering the baseline wavelength range from 0.55 $\mu$m to 11.0 $\mu$m. An optional long wavelength channel extends the range to the goal wavelength of 16.0 $\mu$m, while the baseline design of the visible and near infrared channel \cite{Adriani_1}, \cite{Adriani_2} includes the goal wavelength extension to 0.4 $\mu$m. A Fine Guidance Sensor (FGS), necessary to provide closed-loop feedback to the high stability spacecraft pointing system (AOCS, Attitude and Orbit Control System), is also included in the EChO payload instrumentation (see Fig. ~\ref{fig:1}).

The spectrometer modules share a common field of view, with the spectral division achieved using a dichroic chain operating in long-pass mode. The five core science channels include a cross-dispersed spectrometer (Visible/Near Infrared channel, VNIR) operating in the range from 0.4 to $\sim$2.5 $\mu$m, a grism spectrometer (Short Wave InfraRed channel, SWIR) operating between 2.5 and 5.3 $\mu$m, a prism spectrometer (2 Mid Wave InfraRed channels, MWIR) covering the long wavelength range from 5.3 up to 11 $\mu$m, and a long (optional) wavelength spectrometer operating between 11 and 16 $\mu$m (Long Wave InfraRed channel, LWIR). All science modules and the FGS are accommodated on a common Instrument Optical Bench. The payload instrumentation and the detectors proximity electronics are passively cooled at $\sim$45K with a dedicated instrument radiator for cooling the FGS, VNIR and SWIR detectors to 40 K. An Active Cooler System based on a Neon Joule-Thomson Cooler provides the additional cooling to $\sim$28 K which is required for the longer wavelengths channels (MWIR, LWIR).

The overall EChO payload design, as well as its electrical architecture, has been conceived to maintain a high degree of modularity. This approach helps both technically and programmatically in allowing for independent development of the different modules.

In this paper we provide an overview of the EChO payload electronics and onboard software. The presented architectures are a result of the trade-off processes driven by the mission science requirements, aimed at satisfying both the need to minimize the overall system complexity and the requirements on the overall system budgets.

\section{Payload electrical architecture overview}
\label{payload}

The EChO electrical architecture  \cite{Focardi_1},  \cite{Focardi_2} has been logically divided into two main segments: spectrometerÕs detectors and proximity (cold) electronics on one side, and a spectrometers data handling warm electronics on the other side.

The cold segment hosts the Focal Plane Arrays (FPAs), the ROICs (Read Out Integrated Circuits), the active thermal stabilisation electronics and the cold front-end electronics (CFEEs). The warm electronics contain the warm front-end electronics (FEEs) and implements the unique payload instruments interface to the spacecraft On-Board Computer, hosting nominal and redundant connections for all the digital data links (telemetry, telecommands and housekeeping). The unit implementing these functionalities is called Instrument Control Unit (ICU).

\begin{figure}[!h]
\begin{center}
 \includegraphics[width=10cm]{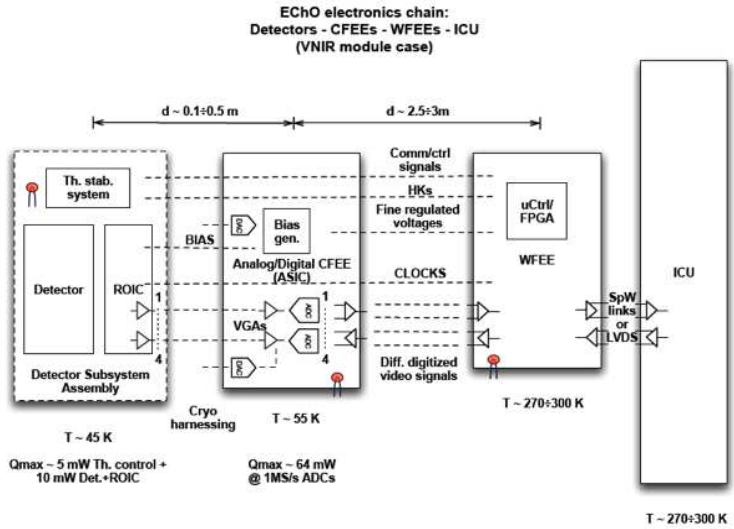}
\caption{EChO VNIR module electrical architecture.}
\label{fig:2}       
\end{center}
\end{figure}

The baseline scientific focal plane arrays are MCT (Mercury-Cadmium-Telluride) detectors for the short wavelengths channels and Si:As for the (optional) longest wavelengths channel. A block diagram of a spectrometer module electrical architecture (e.g. VNIR channel) is shown in Fig. ~\ref{fig:2}. The detectors signals are readout by the Analog/Digital CFEEs ASICs, operating at a temperature of $\sim$50$\div$55K and located at a distance $\le$ 0.5 m from the focal plane arrays. The digital signals are then handled by the warm front end electronics (WFEEs), which are located at a $\sim$2.5$\div$3.5 m distance from the FPA, and sent to the Instrument Control Unit.

The ICU provides the main functionalities to manage all the instrument subsystems implementing detectors commanding, science and housekeeping data acquisition, calibration sources and mechanisms handling and the overall on-board communications management. In the present baseline design WFEEs support the ICU in performing the distributed digital pre-processing of data coming from the CFEEs and in generating their control digital signals. ICU and WFEEs, being warm electronics units operating at $\sim$273$\div$300 K, shall be located inside the SVM (Service Vehicle Module), a module thermally decoupled from the telescope optical bench.

The ICU electronics will rely on a cold-strapped redundant architecture with some trade-off solutions removing or mitigating any electronics single-point failure. Presently the full cold redundancy is foreseen at board level exploiting the ICU passive back plane as the main support for the cross strap routing. This baseline solution will be studied in more detail in the following phase taking into account FMECA (Failure Mode, Effects, and Criticality Analysis) and FDIR (Fault Detection Isolation and Recovery) studies at electronics subsystems level. As alternative solution the full cold redundancy could be implemented duplicating all the electronic boards in a cold-spare cross-strapped configuration.

\subsection{Detectors}
\label{DET}

Within the Consortium, detectors development and testing activities are presently aimed at increasing the technology readiness level of the European candidate MCT detectors operating in the IR range (e.g. arrays of 512 x 512 pixels with 15 $\mu$m pixel pitch coupled to an ASIC for CFEEs from the Dutch SRON) and at developing MWIR new devices. In the present baseline, however, the detectors for the VNIR, SWIR and FGS modules will be Teledyne HgCdTe arrays with 512 x 512 - 18 $\mu$m pixel pitch (H2RG devices \cite{Hale},  \cite{Teledyne_1}, \cite{Bai_1}, \cite{MacDougal}). These arrays are produced in US and are presently the only ones at the required TRL level with an expected noise well within the EChO requirements.

\begin{figure}[!h]
\begin{center}
 \includegraphics[width=6cm]{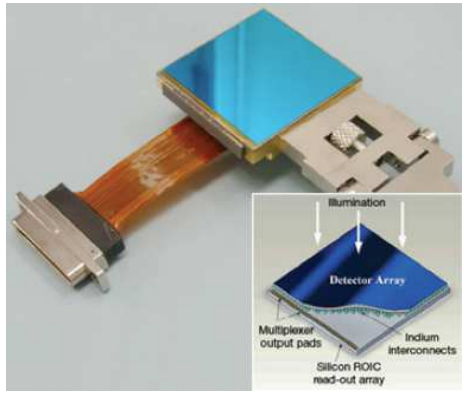}
\caption{An example of a MCT detector from Teledyne: a H2RG hybrid array with 18 $\mu$m pixel pitch. Courtesy Teledyne.}
\label{fig:3}       
\end{center}
\end{figure}

One key outcome during the Assessment Phase was the identification of an MCT detector with TRL approaching 5 that worked up to 11 $\mu$m at an operating temperature of $\sim$40 K.  Due to the lack of a MCT European solution for the MWIR channel, the Teledyne NEOCam device was selected as the baseline. This led to a change in the instrument baseline with the expectation that we could have an all-MCT solution with operating temperatures no lower than 28 K (a constraint due to the desired lower dark current at longest wavelengths). The advantage of this solution is the simplification of the payload cooler by avoiding the use of an active two-stage cooler otherwise required to operate the currently foreseen Raytheon Si:As detectors down to 7 K.

Following these considerations, in the baseline payload design, all spectrometer modules will host MCT-type detectors (Fig. ~\ref{fig:3}) from Teledyne electrically and mechanically bonded with their ROICs.

All ROICs will have at least four or more analogue outputs, amplified and interfaced with their own CFEE (i.e. the SIDECAR ASIC \cite{Beletic_1}, \cite{Beletic_2}) where analogue to digital conversion will take place.

\subsection{Cold Front End Electronics (CFEES)}
\label{CFEES}

\begin{figure}[!h]
\begin{center}
 \includegraphics[width=10cm]{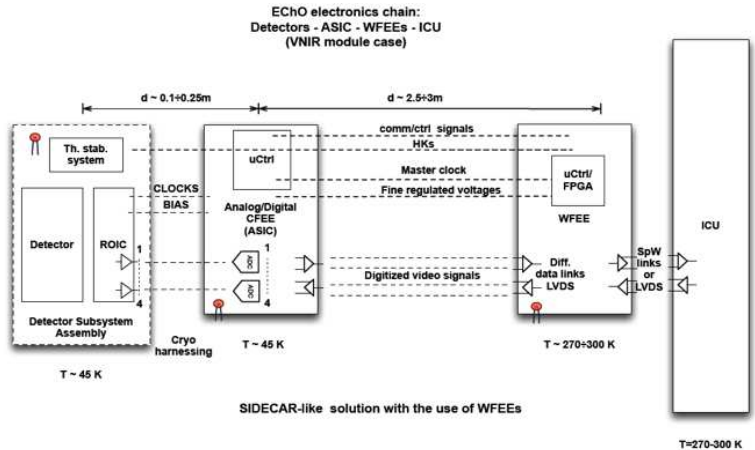}
\caption{EChO payload overall electrical architecture (baseline solution with SIDECAR ASIC and WFEE).}
\label{fig:4}       
\end{center}
\end{figure}

\paragraph{Baseline configuration.} 
As described above, the foreseen baseline solution adopts SIDECAR ASICs from Teledyne as CFEEs for the VNIR, SWIR, MWIR and FGS modules. This solution is the best one to drive properly the US MCT detectors and to save mass, volume and power at the same time. To greatly improve Sensor Chip Arrays (SCAs) integration by reducing size, overall system performance, Teledyne developed the SIDECAR (System for Image Digitisation, Enhancement, Control And Retrieval) ASIC  \cite{Dorn},  \cite{Teledyne_2}. SIDECAR is a SCA interface chip that provides clocks and biases to the SCA, and performs amplification and analog-to-digital conversion of the SCA analog outputs. It can run with up to 36 ADC with a maximum sampling frequency of 500 kHz.
In Fig. ~\ref{fig:4} the ICU is interfaced to the SIDECAR ASICs by means of WFEEs, one per each scientific focal plane array (spectrometer), which is connected to the CFEEs and detectors modules by means of suitable cryo-harness as described in the harnessing section of this paper.

SIDECAR standard version has 36 analog input channels, and digitisation can be done with 12 or 16 bit resolution. SIDECAR presents all-digital interface to the external world, providing the following advantages to system design:

\begin{itemize}
\renewcommand{\labelitemi}{$\bullet$}
\item Simplifies the overall system architecture and reduces the number of wires within the system;
\item Eliminates the risks and problems of transmitting low-noise analog signals over long distances;
\item Offers high flexibility and redundancy due to broad programmability;
\item Digitises 16-bit data at 500 kHz (per port) or 12-bit data at up to 10 MHz (per port) for a maximum total of 12-bit pixel rate at 160 MHz.
\end{itemize}

\begin{figure}[!h]
\begin{center}
 \includegraphics[width=10cm]{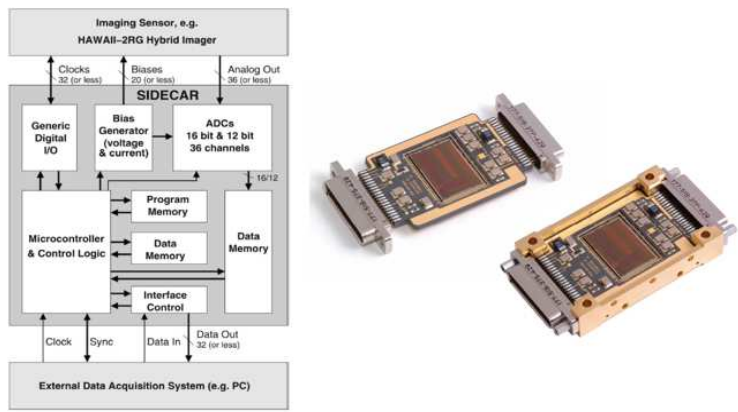}
\caption{Left: SIDECAR electronics block diagram. Right: JWST flight package: ceramic circuit board with ASIC die and passive components, shown without and with metal housing. The dimensions are about 74 x 38 x 10.4 mm$^{3}$. Courtesy Teledyne.}
\label{fig:5}       
\end{center}
\end{figure}

Fig. ~\ref{fig:5} shows a schematic electrical diagram of the SIDECAR ASIC and its interfaces with the H2RG detector (i.e. ROIC multiplexer) and the host (external) electronics. A new version of SIDECAR ASIC has been recently designed by Teledyne \cite{Beletic_4}. The EChO Consortium will evaluate this new product during Phase B in both its I/F configurations with 37 pins microD connector and 91 pins nanoD Airborn connector, if available.

The overall power consumption of a system composed by a H2RG detector readout by a SIDECAR ASIC by means of four 16 bit ADCs (4 channels) at a frequency of 100 kHz is measured to be less than 10 mW \cite{Loose_1}, \cite{Loose_2}.
The detectors FPAs + CFEEs overall power dissipation allocation is a function of wavelength in order to limit the total thermal load from cold module to $< $ 75$\div$80 mW for the VNIR, $< $ 25$\div$30 mW for the SWIR and $< $ 15$\div$20 mW for the MWIR one. These values will include dissipation from the single detector and its thermal stabilisation electronics, CFEE, parasitic loads (conductive and radiative) and all other loads caused directly by the cold-part modules.
An alternative design for the payload electronics has been proposed in which the SIDECAR ASICs are directly interfaced the ICU, negating the need for the WFEEs. This architectural change will be studied in more detail in EChO project Definition Phase (Phase B) and finalised before the SRR (System Requirements Review).

\begin{figure}
\begin{center}
 \includegraphics[width=6cm]{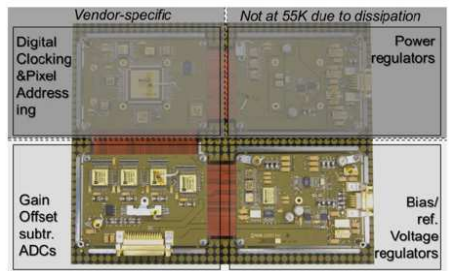}
\caption{Functions implemented in the SRON ASIC are Gain offset subtraction and A/D conversion as well as bias and voltage regulation. Clocking, addressing and power regulation will not be done by this unit but inside the WFEE FPGA (courtesy SRON).}
\label{fig:6}       
\end{center}
\end{figure}

\paragraph{Alternative configuration.} 
The CFEEs alternative solution adopts the SRON CFEE ASIC (Fig. ~\ref{fig:6}). It host up to 4 A/D converters at 16 bits outputs with input VGA (Variable Gain Amplifiers) able to perform offset and gain corrections on the video input signal and at least an externally DAC-controlled fine bias generator, as baseline, to feed the detectors ROICs. A/D converters outputs are differentially interfaced (LVDS) to WFEEs to remove the common mode residual noise.

In order to prevent from and minimising all the likely conductive thermal leakages between the detectors and the CFEEs stages operating at a $\Delta$T of about 10$\div$15 K, the FPAs and the CFEEs shall be electrically connected by harnessing ranging from few cm to about 50 cm. The increasing distance between units could represent a potential issue for pixels clocking and video signal driving from the detectors outputs to the CFEEs, likely requiring an intermediate current buffering stage between the two units, especially when operating the detector at high readout frequencies. This solution would also increment the overall module electrical and electronics complexity leading to a revised basic solution adopting more than two outputs to operate the detectors (e.g. the VNIR channel one) in a low speed mode (below 500 kpx/s @ 16 bits/px) to limit power consumption/dissipation and electrical noise. The final adopted solution with the SRON ASIC should be a compromise between available budgets and the chosen sampling frequency to meet the scientific requirements.

\subsection{Fine Guidance System (FGS)}
\label{FGS}

The main task of the FGS is to ensure the centering, focusing and guiding of the satellite, but it will also provide high precision astrometry and photometry of the target for complementary science. In particular, the data from the FGS will be used for de-trending to aid in the data analysis on ground \cite{Ottensamer}.

The FGS optical module is designed for a 20 arcsec square field of view using 50\% of the flux of the target star below 1.0 $\mu$m wavelength. The optical module provides for internal cold redundancy of the detector chain through the use of an internal 50\%/50\% beam-splitter and two independent detector channels with their own cold and warm drive electronics.

A consortium designed FGS Control Electronics (FCE) unit provides the control and processing of the FGS data and deliver centroid information to the S/C AOCS. The assessment of the FGS accuracy for the faintest target goal star defined for EChO (considering the effective collecting area of the telescope, efficiency parameters of optical elements, beam splitter and QE of the detector) lead to a photoelectrons count of more than 10$^4$ per second. Combined with a pixel scale of 0.1 arcsec and an FWHM of 2$\div$3 pixels, the centroiding accuracy will be less than 0.1 pixel or better than 10 milliarcsecs, well in line with the required precision (PDE - Pointing Drift Error of 20 milliarcsecs -1$\sigma$- over a period of 90 s to 10 hours).

\subsection{Warm Front End Electronics (WFEEs)}
\label{WFEEs}

WFEEs are equipped with digital LVDS transceivers to communicate with CFEEs (to send/receive digital commands and collect digitised data) and host a digital logic (mainly an FPGA and/or a microcontroller) to produce command and control signals and digital clocks to manage the detectors ROICs and CFEEs (mainly the Master Clock and Sync signals for the SIDECAR ASICs).

\begin{figure}[!h]
\begin{center}
 \includegraphics[width=6cm]{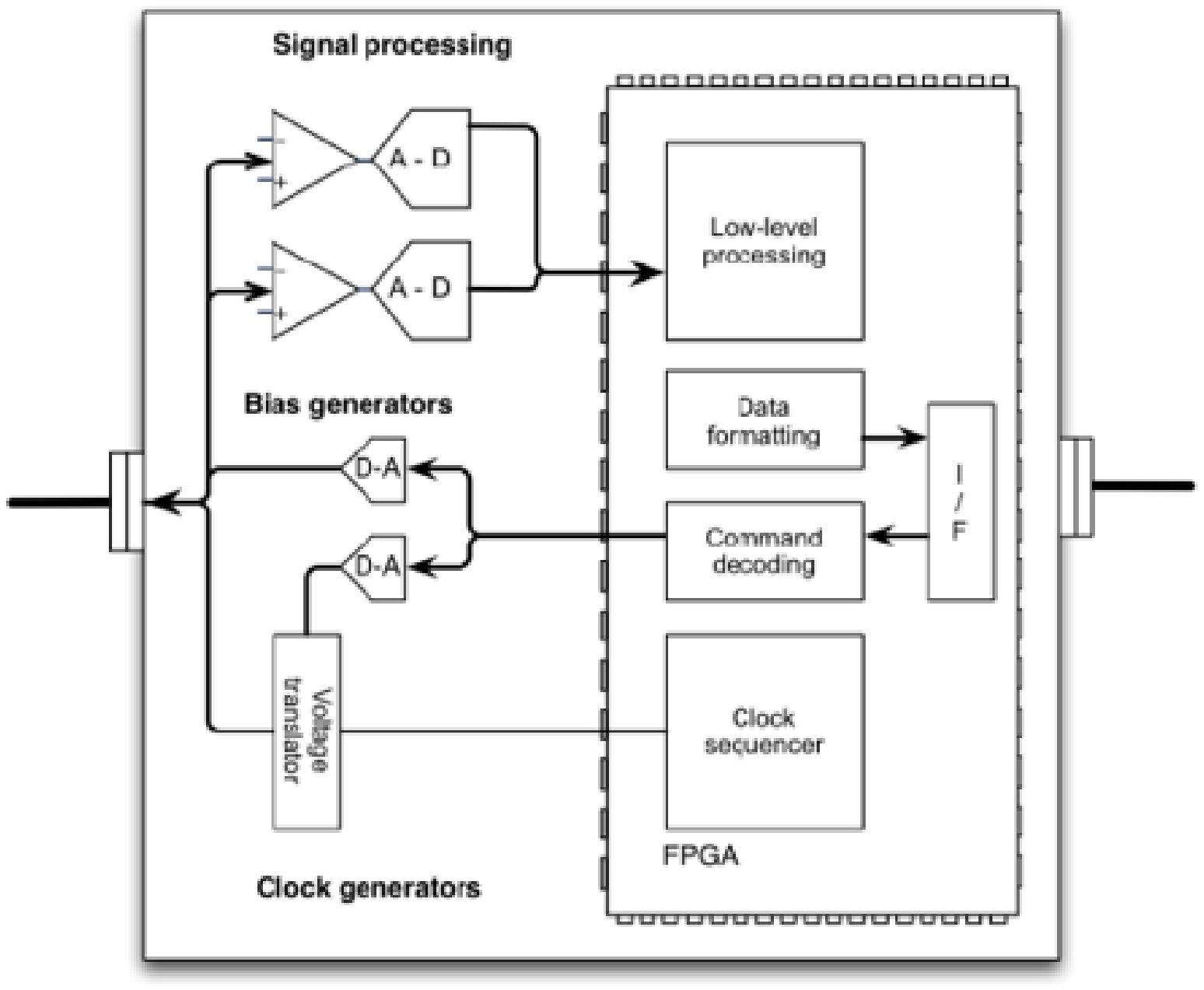}
\caption{Warm Front End electronics block diagram. The scheme is referred to one channel only.}
\label{fig:7}       
\end{center}
\end{figure}

In the baseline design A/D conversion will be performed on board the CFEEs ASIC as well as detector clocking, wave shaping and filtering. The acquired spectra pre-processing (e.g. digital masking and image cropping) will be implemented on board the WFEEs by means of a suitable FPGA (Fig. ~\ref{fig:7}), as a support to the ICU processing.

WFEEs also generate precise secondary voltage levels to feed the CFEEs electronics, which, in turn, produce fine-regulated biases for the detectors FPA assemblies. Also analog housekeepings A/D conversion is performed on board WFEEs before sending digitised values to the ICU, where monitoring and packetisation will take place.

\begin{figure}[!h]
\begin{center}
 \includegraphics[width=6cm]{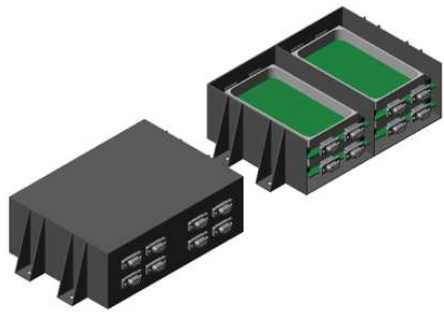}
\caption{WFEEs mechanical design (the number of connectors and their location are only indicative and will be better defined and fixed in the next mission phases).}
\label{fig:8}       
\end{center}
\end{figure}

WFEEs shall be connected to the ICU by means of a nominal plus a redundant SpW link or a LVDS I/F. These interfaces will be implemented by means of IP cores inside the FPGA (e.g. RTAX1000 rad-tolerant device from Actel). The preliminary design for the WFEE electronics boards (FGS excluded) mechanical accommodation is represented in Fig. ~\ref{fig:8}. The aluminium alloy box dimensions are 224x164x80 mm$^{3}$.

\begin{figure}
\begin{center}
 \includegraphics[width=8cm]{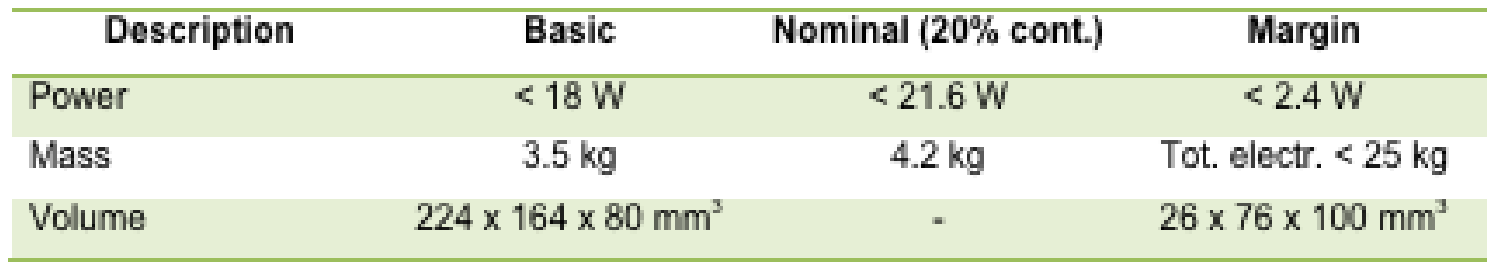}
\caption{WFEEs box mass and volume budgets (for four WFEEs units) with an estimate of the power consumption.}
\label{fig:9}       
\end{center}
\end{figure}

The preliminary WFEEs box mass and volume budgets (for four WFEEs units) are reported in Fig. ~\ref{fig:9} with an estimate of the power consumption. The provided overall dimensions are not taking into account the box mounting feet.

\section{The Instrument Control Unit}
\label{ICU}

The EChO payload electrical architecture is designed to simplify the interfaces between the integrated spectrometer and the spacecraft and to suitably perform the required on-board scientific-data digital processing and proper instrument management.
As described in the previous section, a single Instrument Control Unit is foreseen, implementing all payload instruments control functions and all detectors digital signals acquisition and processing.

\subsection{ICU architecture}
\label{ICU-arch}

\begin{figure}[!h]
\begin{center}
 \includegraphics[width=10cm]{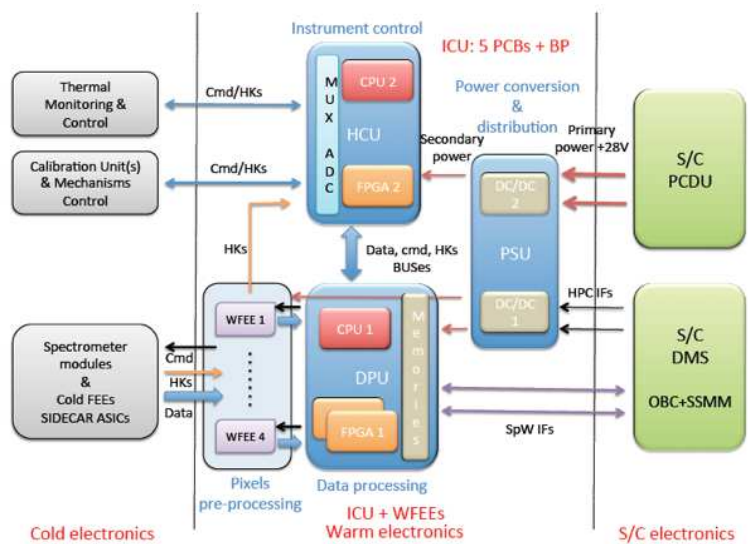}
\caption{EChO ICU architecture block diagram (baseline solution).}
\label{fig:10}       
\end{center}
\end{figure}

In the EChO ICU block diagram reported in Fig. ~\ref{fig:10} the main ICU functional blocks and their interfaces are shown.

The cold proximity electronics are interfaced to the detectors ROICs bonded to the MCT sensors arrays in the VNIR, SWIR, MWIR and LWIR modules (the latter hosts a Si:As detector as baseline). These proximities and the detectors thermal control modules shall provide the buffered analogue HKs (temperatures, voltages and currents) to the warm front-end electronics and ICU, where multiplexing and HKs analogue to digital conversion will take place.

The proposed payload architecture has been designed to minimise and simplify the number of interfaces. In particular, the ICU will be the main payload interface with the spacecraft Power Conditioning and Distribution Unit (PCDU) and the On-Board Computer (OBC) and Solid State Mass Memory (SSMM) Ð (ref. EChO EID-A ESA document).

ICU is mainly a digital unit with processing and data buffering capabilities interfacing both the spacecraft electronics and the digital section of the payload warm front-end electronics for commanding, data and housekeepings (HKs) acquisition, A/D conversion and dispatching, data packetisation and spectra pre-processing, as well as providing finely regulated voltage levels to all its subsystems and FEEs. 

Cold front-end electronics ASICs send digitised data and analog HKs to WFEEs for packetisation in CCSDS format before sending them to the ICU by means of N+R Spacewire (SpW) or, alternatively, LVDS links.
A Data Processing Unit (DPU), a HK and Calibration source management Unit (HCU) and a Power Supply Unit (PSU) are the main blocks of the ICU. 
The currently adopted ICU architecture interfaces 3+1 warm front-end electronics (1 optional module for LWIR) communicating with an on-board ICU digital subsystem with processing capabilities, the Data Processing Unit (DPU) and the Housekeeping and Calibration source management Unit (HCU) both based (as baseline) on a rad-hard space qualified processor and devoted, respectively, to the Data Processing Function and to the Instrument Control Function. Both processors (CPUs) run the main Application SW suitably designed to perform separately the two functions, to manage the payload operating modes (instrument managing) and the overall data acquisition procedures and processing.
The modularity of this design, in which the data processing and instrument control sub-units are well separated will allow for the possibility to have different providers for the sub-units and will ease the integration and test activities at ICU level.

\paragraph{DPU and HCU.} 
The data processing function at pixel level is supported by two rad-hard FPGAs working @ 80 MHz (assumed frequency for the processing resources evaluation) acting as co-processors to perform on-board (HW implemented functions) pixel-based data processing procedures.

The DPU will implement the digital data processing, the data storage and packetisation, the telemetry and telecommand packets handling and the clock/synchronisation capability to temporally correlate the scientific data coming from the different spectrometer channels.

As baseline, DPU is composed by a processing board hosting the scientific digital data processing CPU and the two ancillary FPGAs working as a co-processors and a memory board (or section) hosting all the memories needed to load and run the Application SW (PROMs, EEPROMs and SRAMs) and buffer the incoming scientific data (SDRAMs or FLASH memories) prior digital processing and data formatting. Memories addressing functions are performed by one of the two FPGAs that also act as a memory controller for data buffering.

DPU and HCU will host and share an internal PCI interface to the PCI bus including data, address and control buses. Alternatively an AMBA bus could be used. The PCI bus is connected to the rad-hard processors (an ITAR-free LEON2 AT697F and a LEON3 from ATMEL running at 60 MHz were considered, as baseline, to evaluate the processing and instrument control power needs) and to the main digital logics hosting a PCI I/F (AMBA I/F). 

ICU manages both processors and logics to perform all the required tasks as digital processing (pixels sum, average, digital binning, windowing, spectra cropping and masking, pixels deglitching -if needed-, data and HKs compression) and memory management.

\paragraph{PSU.} 
A Power Supply Unit (PSU) distributes the secondary voltages to the instrument subsystems by means of rad-hard DC/DC converters whereas the HCU will provide instrument/channel thermal control, calibration sources (the common Calibration Unit represented by the integrating sphere on the IOB and the calibration source owning to the VNIR module), mechanisms (the refocusing mirror and the steering mirror if included) and HKs management.

PSU will be feed by two (N+R) +28V power interfaces and will be equipped with two (N+R) BSMs (Bi-Levels Switch Monitors) to report to the S/C the unit status (switched ON/OFF) as defined by two Main (ON/OFF) plus two Redundant (ON/OFF) High Voltage - High Power Pulse Commands (HV-HPC) with interface characteristics as defined by ECSS standards. 

DC/DC converters are hosted by two boards of the PSU in order to feed all the ICU subsystems and provide the basic voltage levels to the WFEEs, so the present ICU design foresees as baseline 5 Extended Double Eurocard (233.4 mm x 220 mm) electronic boards in a fully redundant architecture aboard the same PCBs in a cold strapped configuration exploiting the ICU back panel board interfacing all the electronics subsystems.

\begin{figure}[!h]
\begin{center}
 \includegraphics[width=9cm]{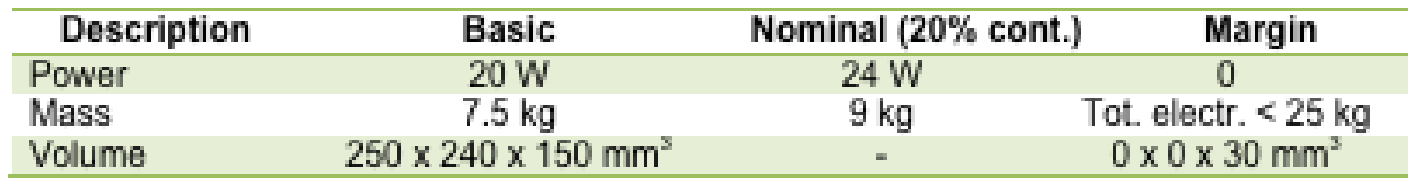}
\caption{ICU power, mass and volume budgets.}
\label{fig:11}       
\end{center}
\end{figure}

 The ICU redundancy policy is based on a trade-off solution removing or reducing to an allowed level the impact of any single-point failure thanks to its cross-strapped fully cold-redundant architecture.  The two DC/DC boards, as baseline, are separately used in a cold-spare redundancy scheme.

The ICU power, mass and volume budgets are reported in Fig. ~\ref{fig:11}. The presented overall dimensions are not considering the box mounting feet.

\subsection{Electrical IFs to the Spacecraft}
\label{IFs-to-SC}

\begin{figure}[!h]
\begin{center}
 \includegraphics[width=5cm]{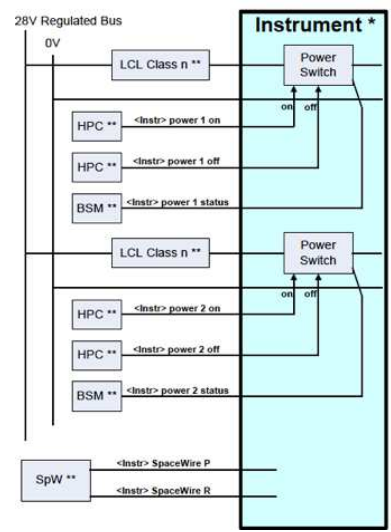}
\caption{Electrical IFs between SVM and the instrument.}
\label{fig:12}       
\end{center}
\end{figure}

\begin{figure}[!h]
\begin{center}
 \includegraphics[width=8cm]{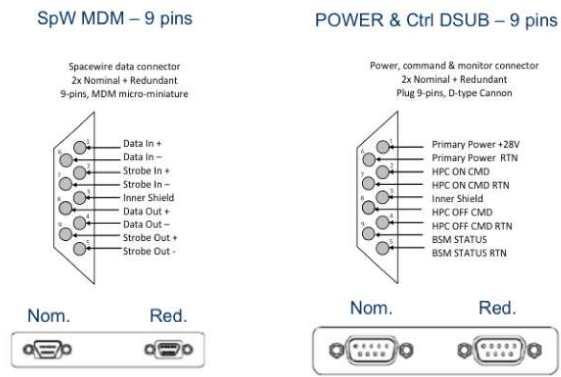}
\caption{TM/TC communications IFs and Power IFs.}
\label{fig:13}       
\end{center}
\end{figure}

The ICU is interfaced to the SVM OBC and SSMM through a +28V power line (Nominal + Redundant), two discrete high voltage - high-power command lines HV-HPC (N+R) for switching on/off the unit and a (N+R) SpaceWire (SpW) interface for commanding, HKs and data transfer to the OBC, configured to work at 10 Mbps (see Figs. ~\ref{fig:12} and ~\ref{fig:13}). The SpW interfaces are implemented inside the DPU service logic FPGA as an IP core that, in turn, is bridged to the PCI bus or the AMBA bus. The same FPGA or a dedicated FPGA controls also the Buffer Memory section used to store data for digital processing and as a temporarily buffer before sending them to the OBC.

In Table 1 the power supply and control lines presently assumed for the EChO payload electronics are indicated. 

Finally, an additional MIL1553 BUS use is under evaluation and could be exploited at payload level e.g. to interface the Fine Guidance System to the S/C.

\begin{table}[!h]
\caption{ICU Power supply and control lines (a cold redundancy is assumed).}
\label{tab:1}       
\begin{tabular}{lll}
\hline\noalign{\smallskip}
Line type & From/To & Nom. and/or Red.  \\
\noalign{\smallskip}\hline\noalign{\smallskip}
Power line: 28V + RTN & (From S/C PCDU to ICU) & N+R \\
Switch On HPC (Signal + RTN) &	(From S/C DMS to ICU) &	N+R\\
Switch Off HPC (Signal + RTN) &	(From S/C DMS to ICU)&	N+R\\
Switch Status BSM (Signal + RTN) &	(From S/C DMS to ICU)&	N+R\\
\noalign{\smallskip}\hline
\end{tabular}
\end{table}

\subsection{Electrical IFs to WFEEs}
\label{IFs-to-WFEEs}

Electrical IFs to the WFEEs are implemented by a nominal and a redundant SpW link (or an alternative LVDS serial link) for every module in order to transmit digital data and HKs telemetry and telecommands. The possibility to use additional analog links to transmit to ICU analog HKs to be multiplexed and digitised by the HCU unit is currently being taken into consideration. Their implementation is subordinated to the final design of the WFEEs and to the final definition of their monitoring needs.

\subsection{Grounding, Isolation, Bonding and Charging Control}
\label{Ground}

\begin{figure}[!h]
\begin{center}
 \includegraphics[width=10cm]{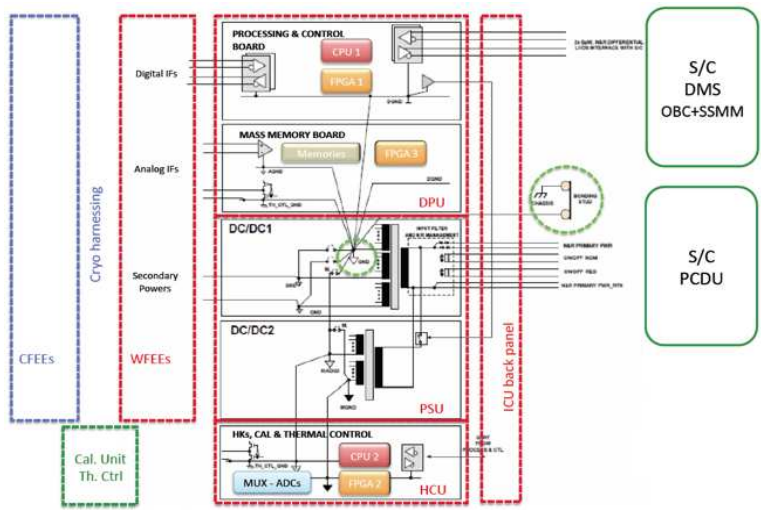}
\caption{ICU grounding concept.}
\label{fig:14}       
\end{center}
\end{figure}

The present EChO ICU grounding concept is a Distributed Single Point Grounding (DSPG), see Fig. ~\ref{fig:14}, involving all the 5 ICU electronics boards. 

The principle is to realise a single point ground for each independent power network and provide isolation between those networks. In order to match this goal, any primary power and return input lead of equipment shall have a DC isolation of at least 1 MOhm (shunted by no more than 50 nF) to equipment chassis and between power input lead and secondary power leads.
In order to avoid grounding loops, secondary power distributed inside the equipment shall be galvanically insulated and decoupled by DC/DC converters.

The ICU box will be bonded to the platform bench via its mounting feet. In addition, the unit shall provide a bonding stud/hole for bonding via a bond strap. All ÒsensibleÓ surfaces will be treated to be electrically conductive, in order to mitigate and control charge accumulation phenomena.

This concept will be revised to guarantee the compatibility with the overall spacecraft grounding scheme, as soon as the spacecraft SVM design will be finalised by the prime contractor.

\subsection{Mechanical design}
\label{Mech}

\begin{figure}[!h]
\begin{center}
 \includegraphics[width=8cm]{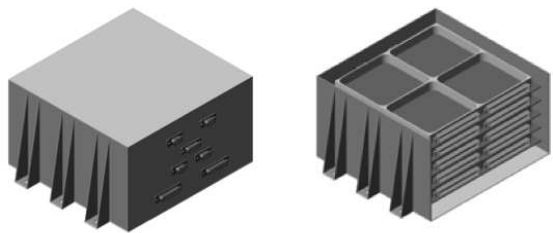}
\caption{ICU mechanical design  (the number of connectors and their location are only indicative and will be better defined and fixed in the next mission phases).}
\label{fig:15}       
\end{center}
\end{figure}

The ICU preliminary mechanical design is presented in Fig. \ref{fig:15}. The ICU box dimensions are 250x240x150 mm$^3$, mounting brackets and feet excluded. All the electronics boards are fixed by means of space qualified card lock retainers and are reinforced by an aluminium structure hosting the connectors as depicted in the figure.

The ICU box, made by aluminium alloy, will have an external bonding stud to connect analog and digital grounds to the unit chassis, in order to implement an alternative configuration for testing and external noise shielding.

\subsection{Thermal design}
\label{Thermal}

The present ICU thermal design is based on a simple radiative and conductive heat transfer analysis performed on similar units, that lead to the overall thermal load dissipation to maintain all the electrical components in their operational and survival ranges and will be refined during the Definition Study (Phase B) of the project. Thermal straps could be foreseen inside the unit to manage properly heat flows between the hottest components and the ICU chassis.

\subsection{Harnessing}
\label{Harness}

ICU IFs harnesses will be divided into EMC classes. All cabling will have an overall screen to provide an EMC shield against radiated emissions and susceptibility and to avoid charging of the cabling dielectric materials. No flying harness is foreseen -as baseline- inside the ICU. No harness dielectric will be directly exposed to space plasma environment.

Cryogenic harnessing from WFEEs to CFEEs is presently under study and its definition requires a close collaboration between the System Team and the Science Team as the speed of acquisition and detectors driving signals have an important impact on the cable design in terms of wire size, electrical properties, shielding configuration etc. All these requirements will then be evaluated in terms of heat leak between thermal stages and, if needed, the design tuned to meet heat load allocations and electrical requirements. At this stage of the study the analysis can be based only on a series of assumptions that shall be reviewed in the next phases of the project \cite{Morgante}.

It is clear from a first quick evaluation that typical standard flat cables or flexi-harness used in such applications (AWG26 37 pins or AWG30 91 pins) based purely on Cu are not a viable solution due to the high parasitic loads that they induce on the stages w.r.t. the allocations on the S/C. Custom made harness in terms of materials, AWG, and shielding, based on standard connectors, shall be required for EChO.

At this stage of analysis the material combination that best suits the EChO set of requirements from both the electrical and thermal point of view are the following:

\begin{itemize}
\renewcommand{\labelitemi}{$\bullet$}
\item CuZn (Brass) for bias control and power supply (the expected voltages/currents should be low enough to avoid, for the moment, Cu wires);
\item PCuSn (Phosphorous Bronze) and/or stainless steel (SS) for analog and digital I/O and all other signals;
\item SS (Stainless Steel) braid for cable shielding.
\end{itemize}

The baseline harnessing between cold and warm electronics is presently considered as a standard round bundle cable configuration (flat cable is assumed only for the connection between Sidecars and detectors and vendor-provided).

\section{Electronics simulators}
\label{Simulator}

The experience gained during the preparation of the Herschel \cite{Liu} and Euclid \cite{Bonoli} satellites instruments has shown the great importance of having subsystem simulators as soon as possible during SW development. The electronics simulators are intended to provide an independent functional platform for testing and simulations. They allow to test the communication protocols, data packetisation, implementation of autonomy functions; without good simulators, all these test activities have to be postponed after the AIV phase with high risk of delay due to the need to correct bugs found in the SW. They can also provide useful data and metrics to analyse and monitor sensors, FEEs electronics and ICU performance.

The present simulator, used as an early testing environment, provides TM and HKs acquisition and is virtually controlling two emulated devices retrieving randomly generated data. 

The development of the simulator is based on a FPGA, which manages all housekeepings, telecommands and telemetry data. All the acquired information is sent to a computer via Ethernet connection for further analysis. The board containing the core of the simulator is GR-XC6S, manufactured by Pender Electronics. The FPGA model is a Xilinx Spartan-6 XC6SLX75.

ICU simulator consists of four modules: one for H/W interfaces and the three remaining ones make up the S/W architecture (simulation, control and monitoring and user interfacing). From this point of view, the system simulation is running the ICU components loaded and provided by the control module, which is also in charge of stand-alone monitoring of every parameter of the system and itÕs operated through the user interface, providing the user with the whole needed functionality for practical operation, configuration and inspection of the simulation system. 

The FPGA integrates an IP core of LEON3 processor. The simulator application running on the development board is coded in C++ and controlled by RTEMS, a fully featured and open-source real-time operating system widely used in embedded platforms.

The software architecture is based on two main components, Device Manager and HK Manager, controlling Device data acquisition and HKs sensor sampling, respectively. Everything is coordinated by a Core component, in charge of TC reception and TM retrieving, and responsible for communication between managers and users.

For the following Definition phase the use of a TSIM LEON3 SW simulator from Aeroflex Gaisler is also foreseen, in order to simulate the main processing tasks of the ICU. This will allow to split the SW and HW functionalities simulations and to cross-check all the results from simulations.

\section{ICU software}
\label{ICU-SW}

The EChO Payload instruments On Board Software (P-OBS) running on the ICU will be composed by the following main blocks:

\begin{itemize}
\renewcommand{\labelitemi}{$\bullet$}
\item \emph{Boot Software}: it is installed in the PROMs of ICU boards and allows for loading the ICU application software (ASW) from EEPROM to RAM at boot. It contains all the low level drivers for the CPU board and its related interfaces.

\item  \emph{Instrument Control Software}: it refers to the software performing the operations of the EChO scientific payload. It implements the data-handling functions of the payload and calibration sources and mechanisms (if present), manages the spectrometers operating modes and runs autonomous function. It implements the interface layer between the S/C and the instruments operation.

\item  \emph{Data processing Software}: implements the ICU science data on-board processing and lossless compression (i.e. RICE). It also implements the data packetisation for the transmission to the S/C Solid State Mass Memory (SSMM).
\end{itemize}

\subsection{EChO On-Board Software functional analysis}
\label{SW-func}

The ICU is responsible for the following main activities:

\begin{enumerate}
\item	Telemetry and Telecommand exchange with the S/C CDMU;
\item	Instrument Commanding, based on the received and interpreted TCs;
\item	Instrument monitoring and control, based on the Housekeeping data (HKs) acquired from the focal plane instrument units;
\item	Synchronisation of all the scientific payload activities;
\item	Detectors readout data acquisition, pre-processing and formatting according to the selected Telemetry protocol;
\item	Science Data download towards the S/C Mass Memory.
\end{enumerate}

\begin{figure}[!h]
\begin{center}
 \includegraphics[width=6cm]{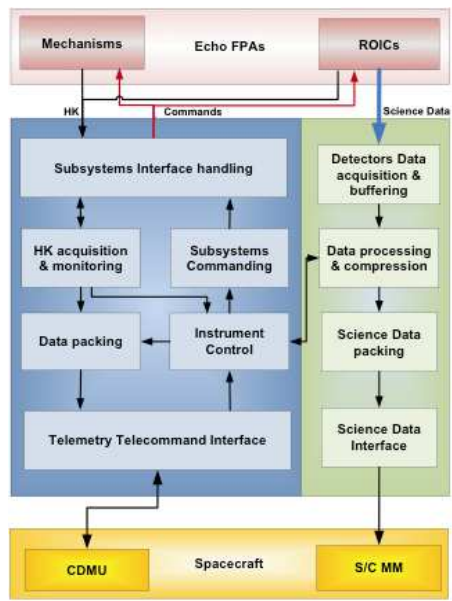}
\caption{ICU OBS functional components block diagram.}
\label{fig:16}       
\end{center}
\end{figure}

These activities can be grouped into the Instrument Control and Data Processing software: these two SWs will constitute the On Board Software (OBS) of the EChO science payload (P-OBS). The On Board Software will be implemented as a real-time multitask application.
The Block Diagram reported in Fig. \ref{fig:16} is a general purpose functional diagram which is applicable to a generic space instrument control and processing software and therefore well represents the main ICU functional architecture. The two different colours indicate the different groups of functionalities of the Instrument Control and the Data Processing software.

\subsection{Software layers}
\label{SW-lay}

\begin{figure}[!h]
\begin{center}
 \includegraphics[width=9cm]{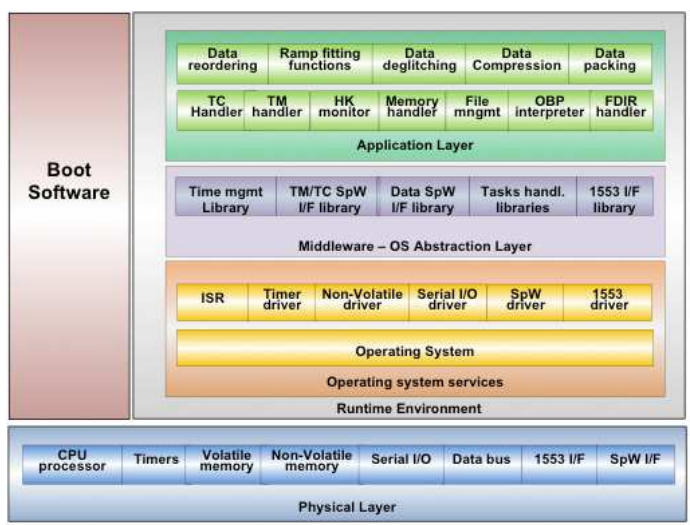}
\caption{SW layers structure of the EChO Payload instruments On Board Software.}
\label{fig:17}       
\end{center}
\end{figure}

The EChO Payload instruments On Board Software layers structure is presented in Fig. \ref{fig:17}. The boot software component refers to the start up software described in the previous section: it is installed in the PROMs of the ICU boards and allows for the application software loading. It contains all the low level drivers for the CPU and its related interfaces.

The physical layer includes all ICU HW components with a direct level of interaction with the on board software. 

The Runtime environment includes the Real Time Operating System (RTOS) layer, necessary to provide multi-tasking support. In case the baseline architecture based on the LEON processor will be confirmed, the RTEMS operating system is a good RTOS candidate, being already used for applications on board ESA satellites. The other indicated operating system services include the drivers for the on-board memories, the on-board HW timers and the local data bus. The other OS services mentioned in the structure of Fig.  \ref{fig:17} are those not directly provided with the OS kernel.
An OS abstraction layer has then been included, in which all middleware libraries have been considered. The middleware services are based on the use of RTOS function calls. They include all library functions dedicated to the low level handling of the ICU HW devices/interfaces. All the middleware libraries will be developed in house and will provide a mean for developing an Application Software virtually independent from the HW and OS below it. This layer is very important and will ease the testing activities. 

The Application Layer includes both the ICU Instrument Control software and the Data Processing software. The ICU Instrument Control SW will implement the functions listed in points from 1 to 4 of the list in sec. ~\ref{SW-func}: i.e. the TM/TC S/C interface handling, the payload housekeeping data acquisition and monitoring, the instruments operating modes management and the autonomous function execution. The present assumption is that the application software will be written in C++, though some functions may need to be coded in assembly to optimise their performance. 

In case stringent timing requirements have to be met for subsystem commanding, an interrupt-driven command sequencer (On Board procedures, OBP interpreter) can be included into the ICU on board software. Based on the experience of HerschelÕs HIFI and SPIRE instrument control software, this is a flexible and effective solution to implement time-critical commanding procedures. Some preliminary tests on the performances of a simplified implementation of such a sequencer on a LEON 3 based SOC implemented on an FPGA development board (GR-XC3S-1500) have been already carried out. A 1 MHz timer has been used to trigger the highest priority interrupt of the system. Under normal workload conditions (only two tasks running, CPU load $<$ 50\%), an average time jitter of the order of 1 $\mu$s (for an overall test duration of more than 1 hour) has been measured, as expected. The proposed sequencer offers high flexibility and re-programmability possibilities, thanks to the straightforward way in which scripts can be reloaded during in-flight operation; this feature can be exploited to modify instrument control or measurement procedures in response to changed mission requirements. The sequencer is described in \cite{DiGiorgio_1} and \cite{DiGiorgio_2}.

\subsection{Boot SW}
\label{Boot}

The boot software is stored in a PROM device, while EEPROM memory devices are used to store two or more images of the Application Software (On Board Software, OBS). The boot software is started automatically after the power on. The boot program is in charge of loading and starting the application program (OBS). The boot process is driven by command packets coming from the Spacecraft. The boot SW will be able to load and start:

\begin{itemize}
\item	one of the OBS images stored in EEPROM (when the Òboot from EEPROMÓ command is received);
\item	an OBS image received by means of specific commands; in particular, a series of memory management commands (Òmemory loadÓ) are used to store in RAM a complete new OBS image, which is then started at the reception of a Òboot from RAMÓ command.
\end{itemize}

The boot procedure described above offers good reliability arguments because:

\begin{enumerate}
\item	If the EEPROM is corrupted, the boot can still load an OBS image via ÒMemory LoadÓ commands.
\item	If any RAM cells in the area to be occupied by the OBS are broken, a new OBS image, which does not use that memory location, can be built on ground and uploaded via telecommands.
\item	Such an approach has been successfully proven during the Herschel in-flight operations.
\end{enumerate}

Of course there still remains the risk of a broken cell in PROM or in the portion of RAM used by the boot SW, which can lead to unrecoverable failures of the boot process. The drawbacks of this procedure are:

\begin{enumerate}
\item	Complex Boot program, it needs to handle the exchange of SpW packets (even if only a very limited set of packets is to be supported);
\item	Big code size, it may be challenging to fit it into PROM device.
\end{enumerate}

\subsection{Data processing}
\label{Data-proc}

\begin{figure}[!h]
\begin{center}
 \includegraphics[width=9cm]{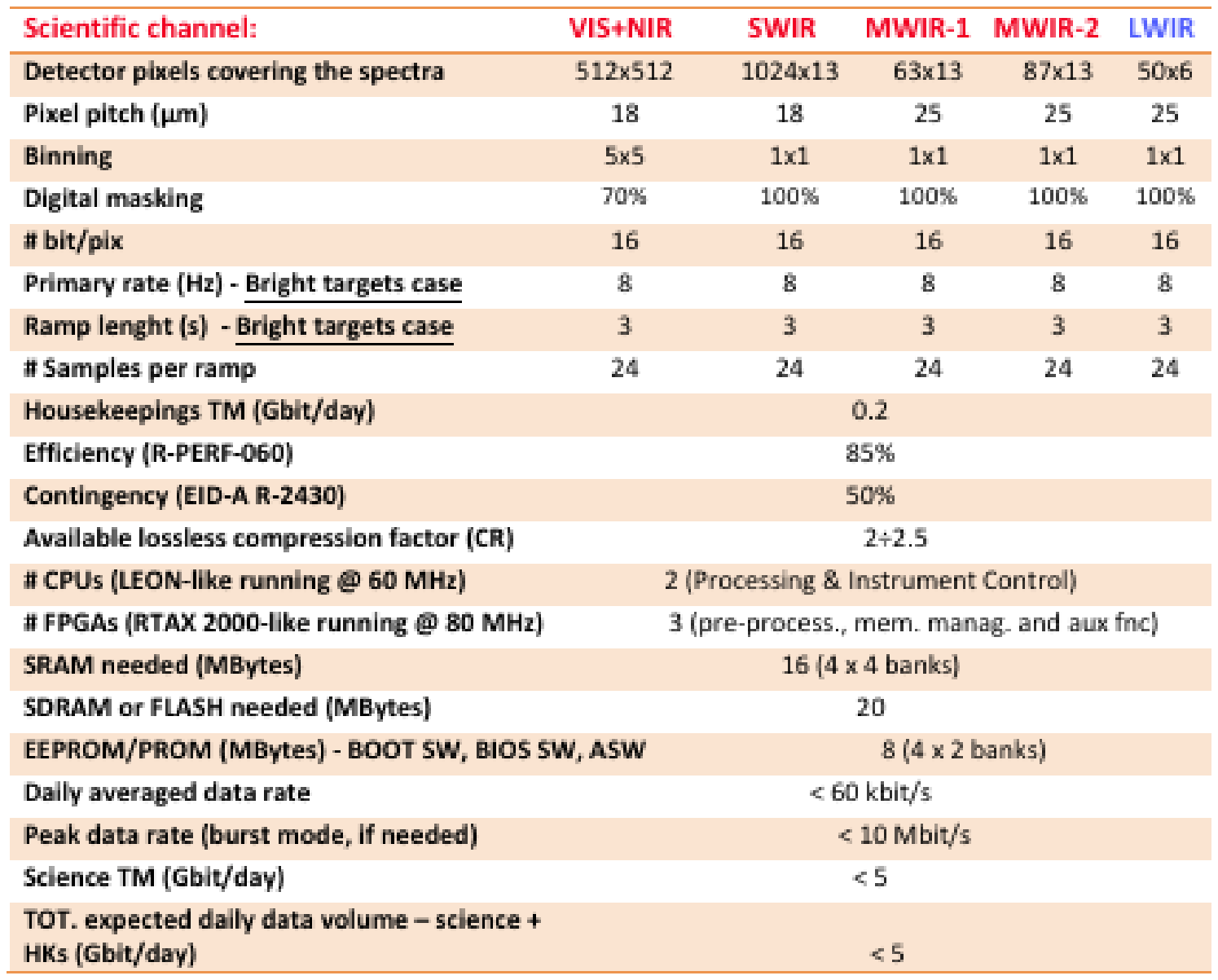}
\caption{ICU processing power, data rate, data volume, required amount of memories.}
\label{fig:18}       
\end{center}
\end{figure}

Fig. \ref{fig:18} reports the expected ICU daily data volume, the data rates to the S/C, the CPUs expected processing power and memories usage. The followings assumptions have been taken into account: the averaged number of clock cycles/elementary operation is 3 (1 MIPS = 3 Mega clock cycles); the number of elementary operations/pixel takes into account the OS background activity overhead; digital data are stored in memory (SRAM and SDRAM/FLASH) before CPU processing; masking is performed by an FPGA aboard WFEEs before sending data to the CPU; payload control refers to TC execution activity between ICU and the (3+1)-modules instrument including calibration units management. Once pre-processed, scientific data are temporarily buffered and then sent to the DMS Solid State Mass Memory.

Our basic assumption is that we sample up-the-ramp pixels in a non-destructive manner with a sampling rate$ \le$ 8 Hz. There will be destructive readouts after different integration times for bright, normal and faint sources, with the length of the ramp determined either by the saturation limits of the detector for a particular target or by the expected maximum length of a ramp, estimated by taking into account the expected cosmic hits rate. This will be determined on-ground by simulations activities and in flight by dedicated calibration procedures. The sampled ramps are fitted and the pixels photocurrent extracted. These data are then processed for pixels cosmic rays deglitching, lossless compressed and transmitted to the S/C in CCSDS format according to the instrumentÕs operating Mode.

The average allocated data rate is 35 Gbits/week or 5 Gbits/day. Assuming an HK production rate of 0.2 Gbits/day, a realistic duty cycle of 85$\div$90\% and a lossless compression ratio of 2, then spending 10\% of the mission in Bright targets mode, 80\% in Normal targets mode and 10\% in Faint targets mode gives a data rate of 4.72 Gbits/day.
All the ICU processing and buffering capabilities are presently evaluated taking into account a 50\% of margin, as usually adopted at the end of the Assessment Study (Phase A) and as required by EID-A.

\emph{Note}: Fig. \ref{fig:18} values strongly depend on data processing required for on-board (ICU) deglitching procedures, compression task and pixels pre-processing. Presently pixels deglitching operation is foreseen on-board but could be avoided given the short integration times for bright targets and the very low likelihood of cosmic rays hits on pixels \cite{Farina}.

\subsection{On-board processing steps}
\label{OBP-steps}

The most demanding module from the point of view of digital processing is the VNIR spectrometer, which is based on a MCT panoramic detector composed by 512 x 512, 18 $\mu$m pixel pitch, on which the VNIR spectra is spread on the two detector halves. Therefore, the main design steps and the system processing needs have been mainly evaluated for this channel. The detectors pixels will be digitally binned (on-chip binning is also possible, as baseline) to produce spaxels, i.e. $NxN$ binned pixels along the spatial and spectral dimensions ($N$=5 in the baseline VNIR design). 

In order to perform on board DPU data processing and pixel deglitching from cosmic rays hitting the VNIR FPA, the two FPA halves (VIS and NIR ones) will be digitally masked and cropped. This pre-processing task will be used to reject all pixels that don't contain any useful information, reducing the overall pixels number to be processed for a factor of the order of 70\% of the entire array. Digital masking and image cropping will be performed aboard WFEEs FPGAs, as baseline, in order to reduce the overall DPU processing load. 

\begin{figure}
\begin{center}
 \includegraphics[width=8cm]{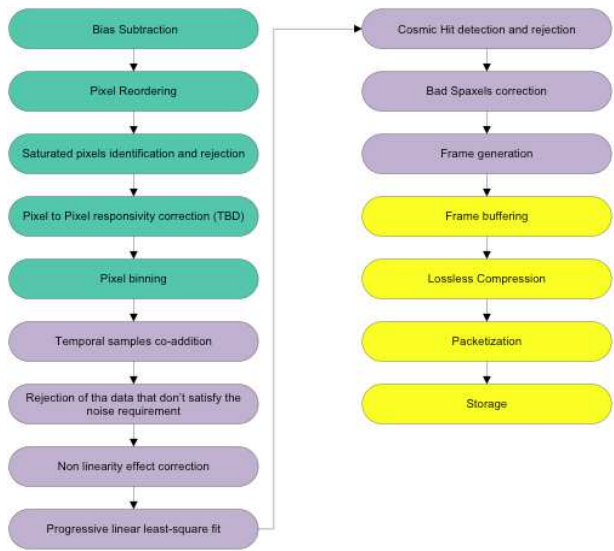}
\caption{Scheme of the data processing steps: green blocks refer to pixel pre-processing activities; purple blocks refer to processing steps at spaxel level and yellow blocks identify frame level processing activities.}
\label{fig:19}       
\end{center}
\end{figure}

The main processing steps are illustrated in Fig. \ref{fig:19}  (mostly for the VNIR channel one, but useful also for all the other channels):

\begin{itemize}
\renewcommand{\labelitemi}{$\bullet$}
\item Bias correction: an internal bias is subtracted to bring all samples to the same ÒgroundÓ. This is necessary for later binning of the data down to spaxel resolution;
\item Pixel reordering: pixels are extracted from the readout electronics output serial stream and reordered, to increase the effectiveness of the subsequent data processing steps. History and ancillary data buffers can be formed at this level;
\item Saturated pixels identification, by defining a cut-off value for the upper limit of the linear region of the detectorÕs response curve. Saturated pixels will be flagged and rejected;
\item Responsivity correction, based on radiometric calibration, using an array responsivity map, will be applied if necessary;
\item Spatial and spectral pixel binning to build spaxels; 
\item Temporal samples co-addition: depending on the ramp fitting algorithm, ÒscansÓ and ÒgroupsÓ have to be formed ($m$ scans per group) if needed;
\item Rejection of the data that donÕt satisfy the noise requirement at the beginning of the ramp, when necessary;
\item Non-linearity correction: if necessary, the ramps will be linearised, even during the multi-accumulate processing);
\item Progressive linear least square fit to calculate ramps slope (pixels/spaxels photocurrents);
\end{itemize}

\emph{Note}: The above last four steps could be necessary to decrease readout noise to specifications. Frames in a group must be contiguous and in the number necessary to reach the required total noise performance limit after co-addition (assuming a multi-accumulation readout mode \cite{Garnett}, \cite{Offenberg_1}, \cite{Raucher}). These processing steps must run at the spaxels readout speed. Using a fixed very simple parameterisation (to be optimised during calibration and algorithm tuning procedures on ground), it can be conveniently performed in hardware.

\begin{itemize}
\renewcommand{\labelitemi}{$\bullet$}
\item 	Cosmic rays and other glitches identification: proper glitch detection and rejection strategies will be implemented, if necessary, with a proper algorithm still to be defined;
\item Bad spaxels correction and linearisation;
\item Frame generation: processed science frames created here will be received on ground;
\item Frame buffering;
\item Lossless Compression (compression of the data to save data volume and bandwidth);
\item Packetisation according to the required TM format: all data products will be packaged;
\item Storage in mass-memory (SSMM) and scheduling for transmission.
\end{itemize}

The listed steps refer to the processing of a single focal plane spectrometer. It is a standard on board processing chain, and a parallel processing of more than one instrument operating simultaneously at the focal plane shall be taken into account. Each detector HK and scientific data owning to a module shall be stored in a different packet store of the S/C SSMM and each payload ASW process shall be identified by a proper APID (Application Process Identifier).

The final overall processing of EChO data will be defined during Phase B and tailored for each one of the focal plane spectrometers. Dedicated simulated data flows will be used to verify the effectiveness of the data reduction steps. In particular the deglitching algorithm performances should be verified against the expected data redundancy, the data acquisition rate and the spaxels dimensions (see next section for more detailed considerations about onboard deglitching needs). Finally, the need to implement lossless compression on-board is strictly related to the results of the on-board deglitching study. If confirmed, a dedicated trade off activity to evaluate the performances of different standard lossless compression algorithms on the on-board CPU processor shall be planned.

\subsection{On-board deglitching}
\label{OB-deg}

Deglitching detectors pixels arrays from cosmic rays hits is commonly an operation to be performed at pixel-level \cite{Fixsen}, \cite{Rausher}, \cite{Offenberg_2}. Operating binning on-chip (e.g. on-board the VNIR detector ROIC) would lead to lose any pixel-based info on cosmic rays hits that would blur the spaxel integrated signal at the same time.
Taking into consideration the predicted cosmic rays flux for the JWST-MIRI instrument at L2, and an integration time from 1.5 to 3 seconds for bright targets, only less than 0.02\% of pixels would be interested by cosmic rays hits, so an option to be considered could be discarding any on-board deglitching operations based on pixels processing at least for bright targets.

The classical deglitching pixel-based procedure is a demanding processing task for the DPU CPU that would require the heavy use of the processor FPU (Floating Point Unit) for the second derivative calculus needed for the glitches recognition and correction. Probably it would be better to perform deglitching at spaxel level (i.e. 25 px/spaxel for the VNIR channel) by detecting cosmic rays hits and discarding them as outliers w.r.t. a median operation performed on sub-arrays. Pixels affected by cosmic rays hits should be flagged and not considered for the rest of the ramp production but properly replaced. A possibility to be explored could be to perform this kind of simplified deglitching procedure at FPGA level (HW-level).

The number of pixels belonging to a spaxel and affected by a cosmic ray hit should be determined by the hit geometry and particles energy. Most likely they would affect the detector array in the direction parallel to the optical axis and the angle defined by the optics F number and baffles, if the detector is properly shielded. So just few pixels should be interested by a single cosmic ray hit and it should be possible to operate deglitching only for the faint targets (if needed) which require longer exposition times with a bigger and not neglecting hit likelihood.

Deglitching from cosmic rays hits has surely a big impact on the overall ICU processing architecture and should be carefully evaluated in respect of the overall system budgets. Deglitching is presently foreseen at CPU level in the ICU processing needs evaluation procedure.

\subsection{Data compression}
\label{Data-comp}

Once performed pixels and spaxels on-board processing all the scientific data related to spectra will be compressed, before sending them to the S/C SSMM. 
Lossless data compression techniques such as zip, gzip, and winzip are widely used for compressing files residing on PCs. These techniques are all based on the Lambel-Ziv-Welch (LZW) \cite{Welch} algorithm or its variations, and would generally yield poor compression ratios on data originating from spacecraft instruments.

A second well-established technique is arithmetic coding \cite{Langdon}. This technique works on most types of data, but exhibits relatively slow speed due to the need to update statistics along the process. The Consultative Committee on Space Data Systems (CCSDS) has adopted the extended-Rice algorithm as the recommendation for international standards for space applications \cite{CCSDS}. This Recommendation defines a source-coding data compression algorithm and specifies how data compressed using the algorithm are inserted into source packets for retrieval and decoding. 

Our present assumption is to implement a SW Rice-like lossless compression. The preliminary tests on very simple simulated spectra lead to lossless compression ratios (CR) well above the needed factor of 2. The processing time and overall performance is strictly dependent on the target CPU, see for example the results of the compression tests on the EUCLID VIS simulated images described in \cite{DiGiorgio_1}.  An activity to assess the feasibility of the lossless compression using a HW compressor is planned for the next mission phase, where the implementation of the compression algorithms at FPGA level (HW-level) in VHDL language or the use of a dedicated ASIC will be evaluated.

\section{Conclusions and future developments}
At the end of the Assessment Phase the baseline EChO Payload electrical architecture has been defined and a preliminary design has been provided. Some still open issues as e.g. the ICU overall processing capabilities depending mainly on the adopted on board deglitching procedures shall be addressed by further analysis. These shall take into account the mission requirements and the available budgets (thermal, mechanical and power dissipation) on a side and the overall system efficiency and reliability on the other side.

Future developments and analysis during the next phase will consider these open issues from the different perspective of the more detailed Definition Phase design.

\begin{acknowledgements}
We acknowledge the financial contribution by the \emph{Italian Space Agency (ASI)} in the framework of the ASI-INAF agreement I/022/12/0 for the  \emph{ÒPartecipazione Italiana allo Studio di Fattibilit\`a della Missione EChOÓ}, the overall partners of the EChO Consortium coordinated by the UK Rutherford Appleton Laboratory (RAL) and the European Space Agency (ESA) for its invaluable support to the development of the Assessment Phase mission design.
\end{acknowledgements}



\end{document}